\newcommand{\op}[1]{{#1}}

\documentclass[osajnl2,superscriptaddress,twocolumn,showpacs,letterpaper]{revtex4}

\usepackage{amsfonts,amssymb,amsbsy,amsmath,bm,graphicx,times,color}

\begin{document}

\title{Nondiffracting beams for vortex tomography}

\author{J. \v{R}eh\'{a}\v{c}ek}
\affiliation{Department of Optics,
Palack\'{y} University, 17. listopadu 12, 
746 01 Olomouc, Czech Republic}

\author{Z. Hradil}
\affiliation{Department of Optics,
Palack\'{y} University, 17. listopadu 12,
746 01 Olomouc, Czech Republic}

\author{Z. Bouchal}
\affiliation{Department of Optics,
Palack\'{y} University, 17. listopadu 12, 
746 01 Olomouc, Czech Republic}

\author{A. B. Klimov}
\affiliation{Departamento de F\'{\i}sica,
Universidad de Guadalajara, 
44420~Guadalajara, Jalisco, Mexico}

\author{I. Rigas}
\affiliation{Departamento de \'Optica, 
Facultad de F\'{\i}sica, Universidad Complutense, 
28040~Madrid, Spain}

\author{L. L. S\'{a}nchez-Soto}
\affiliation{Departamento de \'Optica, 
Facultad de F\'{\i}sica, Universidad Complutense, 
28040~Madrid, Spain}

\begin{abstract}
  We propose a reconstruction of vortex beams based on the
  implementation of quadratic transformations in the orbital angular
  momentum.  The information is encoded in a superposition of
  Bessel-like nondiffracting beams. The measurement of the
  angular probability distribution at different positions
  allows for the reconstruction of the Wigner function.
\end{abstract}

\maketitle

In recent years, considerable attention has been paid to optical
vortices; i.e., beams of light whose phase varies in a 
corkscrew-like manner along the direction of propagation.  
The vortex itself is a topological point defect in the wavefront 
that manifests as a null amplitude because the phase there is
indeterminate~\cite{Soskin:2001yq,Dennis:2009zl}. In addition, these
beams carry orbital angular momentum (OAM) and have a topological
charge $\ell$ that is specified by the total phase change $2 \pi \ell$ 
along a contour around the vortex center.

Apart from their fundamental importance, vortex beams have found
numerous applications~\cite{Franke-Arnold:2008sw}. Besides, entangled
photons prepared in a superposition of states bearing a well-defined
OAM provide access to multidimensional entanglement.  This is of
considerable importance in quantum information and
cryptography~\cite{Mair:2001nv,Molina:2004cc}.

Determining the OAM state of a vortex requires knowledge of
the phase distribution around the singularity.  Since direct
measurement of this phase in the visible is not possible, one needs to
rely on interferometric techniques that work reasonably well both at the
single-photon level~\cite{Leach:2002nn} and for classical
beams~\cite{Harris:1994xy,Berkhout:2008ty}.

The goal of this Letter is to bring forward a feasible \textit{direct}
reconstruction of an optical vortex. We recall that efficient methods
of state reconstruction are of the greatest relevance.  Since the
first theoretical proposals, this discipline has witnessed a
significant growth and laboratory demonstrations of state tomography
span a broad range of systems~\cite{Jarda:2004yk,Lvovsky:2009zk}.

The canonically conjugate variable to OAM is the angle, so we shall
need some basic concepts about this quantity. Through all this Letter
we use a quantum notation, although the translation to a purely
classical language is direct.  Let us denote by $\op{L}$ the OAM along
the direction $z$ of propagation. For our purposes, the simplest
choice is to use the complex exponential of the angle $\op{E} = e^{- i
  \op{\phi}}$, which satisfies the commutation relation $[ \op{E},
\op{L} ] = \op{E}$~\cite{Rehacek:2008ss}.  The action of $\op{E}$ on
the OAM eigenstates is $\op{E} | \ell \rangle = | \ell - 1 \rangle$,
and it possesses then a simple implementation by means of phase mask
removing a charge $+ 1$ from a vortex state~\cite{Hradil:2006bc}.
Since the integer $ \ell $ runs from $ - \infty$ to $+ \infty$,
$\op{E}$ is a unitary operator whose eigenvectors $ | \phi \rangle =
\frac{1}{\sqrt{2\pi}} \sum_{\ell \in \mathbb{Z}} e^{- i \ell \phi} |
\ell \rangle$ are states with a well-defined angle.  A point particle
is necessarily located at a single value of the periodic angular
coordinate. However, for extended objects, such as fields, the
eigenvalue $\phi$ might be not directly related to an angular position
in the transverse plane, as we shall see.

This canonical pair describes the physics of a planar rotor.  The
phase space is then the discrete cylinder $\mathcal{S}_{1} \times
\mathbb{Z}$, where $\mathcal{S}_{1}$ is the unit circle (associated to
the angle) and the integers $\mathbb{Z}$ translate the discreteness of
the OAM~\cite{saverio}. To properly represent a state described by the
density matrix $\varrho$, we need to introduce a Wigner
function. Following our approach in Ref.~\cite{Rigas:2008ac}, we
define the Wigner function for this pair as
\begin{equation}
  \label{eq:WignerAngle}
  W (\ell, \phi) = \frac{1}{2\pi} \int_{-\pi}^{\pi} d\phi^{\prime} \, 
  e^{i \ell \phi^{\prime}} \, 
  \langle \phi -  \phi^{\prime}/2 | \op{\varrho} |
  \phi + \phi^{\prime}/2 \rangle  \, ,
\end{equation}
whose analogy with the standard one for position and momentum is more
than evident.  It is advantageous to rewrite
Eq.~(\ref{eq:WignerAngle}) as
\begin{equation}
  \label{eq:Wig11}
  W (\ell, \phi) = \frac{1}{2\pi} \sum_{\ell^{\prime} \in \mathbb{Z}}
  \int_{-\pi}^{\pi} d\phi^\prime  \,  
  e^{i(\ell^\prime \phi  - \ell \phi^\prime)} \,
  \varrho (\ell^\prime, \phi^\prime)  \, , 
\end{equation}
where $\varrho (\ell^\prime, \phi^\prime)$ are the Fourier
coefficients of $W (\ell, \phi)$, given by $\varrho (\ell,\phi) = 
 e^{- i\ell \phi/2} \sum_{\ell^\prime} e^{i \ell^{\prime} \phi} \, \langle
\ell^\prime | \op{\varrho} | \ell^\prime - \ell \rangle$.  For $\ell =
0$ this is simply the Fourier transform of the OAM spectrum $\langle
\ell| \,\op{\varrho}\,|\ell\rangle$.

As one step further, assume we are able to measure $\op{L}^2$
transformations on the input state followed by angular projections,
that is,
\begin{equation}
  \label{eq:L2}
  \omega (\phi, \zeta ) = \langle \phi | 
  \exp(i \zeta \op{L}^2/2) \, \op{\varrho} \, 
  \exp(- i  \zeta \op{L}^2/2) | \phi \rangle \, . 
\end{equation}
Together with the OAM spectrum, these tomograms $ \omega (\phi, \zeta
)$ provide complete information.  Indeed, one can check that
\begin{equation}
  \label{eq:Tom14}
  \varrho (\ell, \phi) = \frac{1}{2 \pi} 
  \int d\phi^\prime \,  e^{-i \ell \phi^\prime} \,
  \omega (\phi^\prime,\phi/\ell ) \, ,
\end{equation}
so the measurement of $\omega (\phi, \ell)$ allows for the
determination of $\varrho (\ell, \phi)$ and hence the full
reconstruction of the Wigner function via Eq.~\eqref{eq:Wig11}.

The feasibility of this scheme relies on the implementation of the
$\op{L}^{2}$ operation, which is precisely a free rotor.  Our starting
point is a monochromatic light beam whose complex scalar profile can
be written in cylindrical coordinates as $\Psi (r, \varphi, z ) = \psi
( r, \varphi, z ) \, \exp(-i k z)$, where the envelope $\psi ( r,
\varphi, z )$ is a slowly varying function of $z$.  In further
consideration, we restrict ourselves to the paraxial regime and 
neglect polarization effects.

In the angle representation, $\op{L}$ acts as $- i \partial_{\varphi}$
and the OAM eigenstates (which constitute an appropriate basis) are
conveniently expressed (in the plane $z=0$) by
\begin{equation}
  \label{mwave}
  \psi_{\ell} (r, \varphi) = \langle r, \varphi | \ell \rangle = 
  A_{\ell} (r) \exp(i \ell \varphi) \, ,
\end{equation}
where $A_\ell (r)$ is an arbitrary radial profile. Please, notice
carefully that the polar angle $\varphi$ in the transverse plane
should not be confused with the rotor eigenstate $\phi$.  Next, we
look at beams for which free propagation induces an $\op{L}^2$
transformation, as for them the tomography will be enormously facilitated. 
We thus require
\begin{equation}
  \label{evol}
  \psi_{\ell} (r, \varphi, z) = e^{i a \ell^2 z} \, 
  \psi_\ell (r, \varphi) \, .
\end{equation}
In this case the \textit{free propagation} of the optical field
corresponds to \textit{free evolution} of the rotor. Since the free
evolution of paraxial optical beams is determined by the wave
equation $ \Delta_{\perp} \psi_\ell - 2 i k \partial_{z} \psi_{\ell} = 0$, 
where $\Delta_{\perp}$ is the transverse Laplacian,  we obtain 
\begin{equation}
  r^2 \frac{d^2 A_{\ell}(r)}{d r^2} + r \frac{dA_{\ell}(r)}{dr}
  +(\ell^2 r^2 - 2 a k \ell^2) A_{\ell}(r) = 0 \, ,
\end{equation}
whose regular solution is given in terms of the Bessel function
$A_{\ell}(r) \propto J_{\ell} (\sqrt{2 a k} \,\ell r)$.  Consequently,
the beams we look for turn out to have an envelope 
\begin{equation}
\label{vortsp}
 \psi_{\ell} (r, \varphi, z) = N_{\ell} \, J_{\ell} (\sqrt{2 a k}\,\ell r) \, 
  e^{i a \ell^{2} z} \, e^{i \ell \varphi}  \, .
\end{equation}
As it happens with plane waves, these beams require infinite energy,
so their practical generation relies in a spatial apodization of
some kind. We take this into account in the normalization
factor  $N_{\ell}^{2} = [2 \pi \int_{0}^{R} dr \, r \, J_{\ell}
(\sqrt{2 a k}\,\ell r)]^{-1}$, $R$ being the radius of this apodizing 
element.  

Taking $a \sim 1$ guarantees that we are safely in the paraxial region
$|a|\ll 2 k/\ell^2$. In that case, the Bessel beams involved have
central maxima of the size of a few millimeters and quadratic phases
$a \ell^2 \sim 10$ rad/m are induced in the free propagation. This 
confirms the feasibility of our proposal.

We can also understand the form of the vortices (\ref{vortsp}) in an
alternative way.  The paraxial free-space transfer function is $H
(\rho, \theta) \propto \exp ( -i \pi \lambda z \rho^2 )$, where
$(\rho, \theta)$ are polar coordinates in the spatial-frequency space
and $\lambda$ is the wavelength. Let us assume that all the
nonvanishing frequency components of our field lie on a circle of
radius $\rho_0$, that is, $\widehat{\psi} (\rho, \theta) \propto
\widehat{\psi} (\theta) \, \delta (\rho - \rho_0)$, where
$\widehat{\psi} (\rho, \theta) = \mathcal{F} [ \psi(r, \varphi)]$
denotes the Fourier transform. Then, free propagation contributes just
by an unessential overall phase factor, $ \widehat{\psi}^{\prime}
(\rho, \theta) = H (\rho , \theta) \, \widehat{\psi} (\rho, \theta ) =
\exp(-i \pi \lambda z \rho_0^2 ) \, \widehat{\psi} (\theta)$, so the
transverse intensity distribution becomes independent of the distance
$z$ (and hence represents a nondiffracting beam).  The particular case
of the family~\eqref{evol} is obtained by taking $\rho_0 = \sqrt{2 a
  k} \, |\ell|/(2\pi)$.

A general input state can be encoded into an optical beam by creating
a superposition of these basis waves $\psi (r, \varphi ) = \sum_{\ell}
c_{\ell} \, \psi_{\ell} (r, \varphi)$, which can be easily
accomplished by an amplitude spatial light modulator displaying a
hologram computed as an interference pattern of the required field
$\psi (r, \varphi )$ and a reference plane wave.

After propagating a distance $z$ the beam will evolve into a new
superposition $\psi (r, \varphi, z) = \sum_{\ell} c_{\ell} e^{i a  \ell^{2} z} 
\, \psi_{\ell} (r, \varphi)$.  But according to Eq.~(\ref{eq:L2}),
this  is just the required $\op{L}^{2}$ action with $\zeta= -a z$.  
This is the main result of this paper: by chosing a particular form 
\eqref{vortsp} of the OAM eigenstates, the highly
nontrivial $L^2$ transformation is realized by a simple free beam
propagation.  This also provides an experimentally feasible optical
realization of a free rotor.

After some simple calculations, the associated tomograms for the
superposition field turns out to be
\begin{equation}
  \label{eq:tomlh}
  \omega(\phi,\zeta)=  \frac{1}{2 \pi} \left | 
    \sum_{\ell \in \mathbb{Z}} c_\ell \, e^{i\ell \phi} \,
   e^{-i\zeta \ell^2/2} \right |^{2} \, .
\end{equation}
By scanning the values of $\zeta$, the beam can thus be reconstructed
according to Eq.~\eqref{eq:Tom14}. This completes our proposed scheme.
Naturally, \eqref{eq:tomlh} are independent of the rapid oscillating
factor $\exp (- i k z)$.

As a last experimental detail, we need to assess the measurement of
the angular spectrum in Eq.~\eqref{eq:L2}. For the field $\psi (r,
\varphi)$ this reads as (taking for simplicity the plane $z=0$)
\begin{equation}
  \label{averaging}
  \omega (\phi, 0) =  P( \phi ) =
  \left | \int  d\varphi dr \, r \, \psi_{\phi}^\ast (r, \varphi)
    \, \psi (r,\varphi)   \right|^2 \, ,
\end{equation}
where $\psi_{\phi} (r, \varphi) = \langle r,\varphi | \phi \rangle$, 
and $| \phi \rangle$ is a field with well-defined angle.  Note in passing
that the angular spectrum $P(\phi)$ cannot be obtained, in general, 
by measuring the angular profile of the output beam, e.g., using a
wedge-shaped aperture.  In fact, as a consequence of the nontrivial
radial profile of the OAM eigenstates, the states of $\psi_{\phi} (r,
\varphi)$ are given by
\begin{equation}
  \label{anglefield}
  \psi_{\phi} (r, \varphi) = \sum_{\ell \in \mathbb{Z}} 
  e^{i \ell (\varphi-\phi)} \, J_{\ell} (\sqrt{2 a k} \,\ell r) \,  .
\end{equation}

Finally, using the cross-correlation theorem of Fourier analysis, one
gets $ P (\phi) = \left | \mathcal{F}^{-1} [
\widehat{\psi}_{\phi}(\rho,-\theta) \; \widehat{\psi} (\rho,\theta)]
\right|^2_{r=0}$.  This result provides a simple interpretation of the
tomographic setup. The angular states \eqref{anglefield} are defined
in such a way that by placing a transparency $\widehat{\psi}_{\phi}
(\rho,-\theta)$ in the Fourier plane, the unwrapping of azimuthal
phase modulations occurs simultaneously for all the vortices in the
superposition $\psi(r,\varphi)$.  As a result, the on-axis
interference observed at the output plane of the optical processor
yields phase information about the coefficients $c_{\ell}$.  Different
angular components can be readily accessed by rotating the filter
transparency with respect to the optical axis.

The proposed experiment thus consists of the following four steps: (i)
The input state is encoded into a superposition of Bessel-like beams
$\psi_l(r,\varphi)$.  (ii) This superposition is propagated a distance
$z$.  (iii) The angular distribution $\omega(\phi,z)$ is determined by
measuring the intensity in the middle of the output plane of the
optical processor as a function of the Fourier filter orientation
$\phi$.  The last two steps are repeated for a set of distances $z$.
(iv) The Wigner function of the input state is reconstructed with the
help of Eqs.~\eqref{eq:Wig11} and \eqref{eq:Tom14}.

\begin{figure}
  \centerline{\includegraphics[width=0.95\columnwidth]{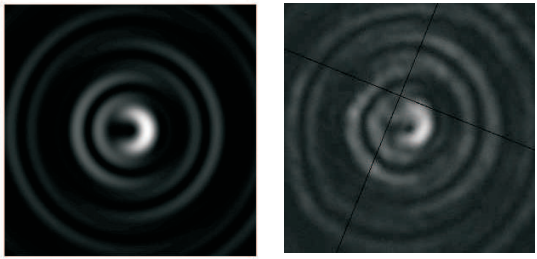}}
 \caption{(Color online) Calculated (left) and experimentally
generated (right)  transverse intensity distribution for a normalized
superposition of the states $\psi_{1} (r, \phi)$ and $\psi_{2} (r, \phi)$.}
\label{fig:comparison}
\end{figure}

As an illustration, we apply our reconstruction procedure to the state
$ \psi (r, \varphi ) = [ \psi_{1} (r, \varphi ) + \psi_{2} (r,
\varphi) ]/ \sqrt{2}$.  The calculated transverse intensity of this
state is shown in the left panel of Fig.~\ref{fig:comparison}, while a
typical CCD scan of the state generated in the laboratory by an
amplitude spatial light modulator (CRL Opto 1024$\times$768 pixels) is
shown on the right for comparison. As an illustration, we apply our
reconstruction procedure to the state $ \psi (r, \varphi ) = [
\psi_{1} (r, \varphi ) + \psi_{2} (r, \varphi) ]/ \sqrt{2}$.  The
calculated transverse intensity of this state is shown in the left
panel of Fig.~\ref{fig:comparison}, while a typical CCD scan of the
state generated in the laboratory by an amplitude spatial light
modulator (CRL Opto 1024$\times$768 pixels) is shown on the right for
comparison.

\begin{figure}[b]
  \centerline{\includegraphics[width=0.90\columnwidth]{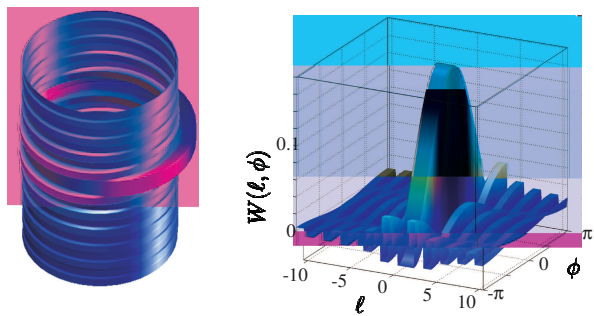}}
  \caption{(Color online) Reconstructed Wigner function
    (\ref{eq:FinalWig}) corresponding to the same state as in Fig.~1.
    Left panel shows this function on the unit cylinder (whose
    vertical axis corresponds to the OAM), while right panel show the
    unwrapped plot in Cartesian axis.}
  \label{fig:tornillo}
\end{figure}

The tomograms are calculated from Eq.~\eqref{eq:tomlh}, whence we
obtain the Fourier coefficients as $\varrho (\ell,\phi) = e^{i  3\phi/2} 
 ( \delta_{\ell, 1} + \delta_{\ell,-1})/(4 \pi)$. This leads to the Wigner function
\begin{equation}
  \label{eq:FinalWig}
  W (\ell,\phi) = \frac{1}{4\pi} (\delta_{\ell, 1} + \delta_{\ell, 2} ) +
  \frac{(-1)^{\ell +1 }}{\pi^2(3- 2 \ell)} \,\cos \phi \, .
\end{equation}
We can see three contributions: two flat slices, coming from the
fields $\psi_{1}$ and $\psi_{2}$, located at $\ell=1$ and $\ell=2$,
respectively, and an interference term that contains contributions
from all the values of $\ell$, although damped as $1/\ell$. These
features are illustrated in Fig.~\ref{fig:tornillo}, which shows the
reconstructed Wigner function on the unit cylinder and its wrapped
counterpart in Cartesian axes.

In practice, one cannot scan infinite values of $\zeta$ (i.e., of
$z$), and only some discrete version of the Eq.~(\ref{eq:Tom14}) is
available in the laboratory. We have checked that with 50 equispaced
values, the reconstruction is still fairly stable.

In summary, we have carried out a full program for a complete
reconstruction of vortex beams. In our scheme, the information is 
encoded in a superposition of Bessel-like nondiffracting beams for
which free propagation is equivalent to the $L^{2}$ operation.
At the same time,  our formulation provides a theoretical 
framework on the action of angle and angular momentum
on these beams.


\begin{thebibliography}{17}
\expandafter\ifx\csname natexlab\endcsname\relax\def\natexlab#1{#1}\fi
\expandafter\ifx\csname bibnamefont\endcsname\relax
  \def\bibnamefont#1{#1}\fi
\expandafter\ifx\csname bibfnamefont\endcsname\relax
  \def\bibfnamefont#1{#1}\fi
\expandafter\ifx\csname citenamefont\endcsname\relax
  \def\citenamefont#1{#1}\fi
\expandafter\ifx\csname url\endcsname\relax
  \def\url#1{\texttt{#1}}\fi
\expandafter\ifx\csname urlprefix\endcsname\relax\def\urlprefix{URL }\fi
\providecommand{\bibinfo}[2]{#2}
\providecommand{\eprint}[2][]{\url{#2}}

\bibitem[{\citenamefont{Soskin and Vasnetsov}(2001)}]{Soskin:2001yq}
\bibinfo{author}{\bibfnamefont{M.}~\bibnamefont{Soskin}} \bibnamefont{and}
  \bibinfo{author}{\bibfnamefont{M.~V.} \bibnamefont{Vasnetsov}},
 \bibinfo{journal}{Prog. Opt.} \textbf{\bibinfo{volume}{41}},
  \bibinfo{pages}{219} (\bibinfo{year}{2001}).


\bibitem[{\citenamefont{Dennis et~al.}(2009)\citenamefont{Dennis, O'Holleran,
  and Padgett}}]{Dennis:2009zl}
\bibinfo{author}{\bibfnamefont{M.~R.} \bibnamefont{Dennis}},
  \bibinfo{author}{\bibfnamefont{K.}~\bibnamefont{O'Holleran}},
  \bibnamefont{and} \bibinfo{author}{\bibfnamefont{M.~J.}
  \bibnamefont{Padgett}},
 \bibinfo{journal}{Prog. Opt.} \textbf{\bibinfo{volume}{53}},
  \bibinfo{pages}{293} (\bibinfo{year}{2009}).


\bibitem[{\citenamefont{Franke-Arnold et~al.}(2008)\citenamefont{Franke-Arnold,
  Allen, and Padgett}}]{Franke-Arnold:2008sw}
\bibinfo{author}{\bibfnamefont{S.}~\bibnamefont{Franke-Arnold}},
  \bibinfo{author}{\bibfnamefont{L.}~\bibnamefont{Allen}}, \bibnamefont{and}
  \bibinfo{author}{\bibfnamefont{M.}~\bibnamefont{Padgett}},
  \bibinfo{journal}{Laser Photon. Rev.} \textbf{\bibinfo{volume}{2}},
  \bibinfo{pages}{299} (\bibinfo{year}{2008}).


\bibitem[{\citenamefont{Mair et~al.}(2001)\citenamefont{Mair, Vaziri, Weihs,
  and Zeilinger}}]{Mair:2001nv}
\bibinfo{author}{\bibfnamefont{A.}~\bibnamefont{Mair}},
  \bibinfo{author}{\bibfnamefont{A.}~\bibnamefont{Vaziri}},
  \bibinfo{author}{\bibfnamefont{G.}~\bibnamefont{Weihs}}, \bibnamefont{and}
  \bibinfo{author}{\bibfnamefont{A.}~\bibnamefont{Zeilinger}},
  \bibinfo{journal}{Nature} \textbf{\bibinfo{volume}{412}},
  \bibinfo{pages}{313} (\bibinfo{year}{2001}).

\bibitem[{\citenamefont{Molina-Terriza
  et~al.}(2004)\citenamefont{Molina-Terriza, Vaziri, Rehacek, Hradil, and
  Zeilinger}}]{Molina:2004cc}
\bibinfo{author}{\bibfnamefont{G.}~\bibnamefont{Molina-Terriza}},
  \bibinfo{author}{\bibfnamefont{A.}~\bibnamefont{Vaziri}},
  \bibinfo{author}{\bibfnamefont{J.}~\bibnamefont{Rehacek}},
  \bibinfo{author}{\bibfnamefont{Z.}~\bibnamefont{Hradil}}, \bibnamefont{and}
  \bibinfo{author}{\bibfnamefont{A.}~\bibnamefont{Zeilinger}},
  \bibinfo{journal}{Phys. Rev. Lett.} \textbf{\bibinfo{volume}{92}},
  \bibinfo{pages}{167903} (\bibinfo{year}{2004}).


\bibitem[{\citenamefont{Leach et~al.}(2002)\citenamefont{Leach, Padgett,
  Barnett, Franke-Arnold, and Courtial}}]{Leach:2002nn}
\bibinfo{author}{\bibfnamefont{J.}~\bibnamefont{Leach}},
  \bibinfo{author}{\bibfnamefont{M.~J.} \bibnamefont{Padgett}},
  \bibinfo{author}{\bibfnamefont{S.~M.} \bibnamefont{Barnett}},
  \bibinfo{author}{\bibfnamefont{S.}~\bibnamefont{Franke-Arnold}},
  \bibnamefont{and} \bibinfo{author}{\bibfnamefont{J.}~\bibnamefont{Courtial}},
  \bibinfo{journal}{Phys. Rev. Lett.} \textbf{\bibinfo{volume}{88}},
  \bibinfo{pages}{257901} (\bibinfo{year}{2002}).

\bibitem[{\citenamefont{Harris et~al.}(1994)\citenamefont{Harris, Hill,
  Tapster, and Vaughan}}]{Harris:1994xy}
\bibinfo{author}{\bibfnamefont{M.}~\bibnamefont{Harris}},
  \bibinfo{author}{\bibfnamefont{C.~A.} \bibnamefont{Hill}},
  \bibinfo{author}{\bibfnamefont{P.~R.} \bibnamefont{Tapster}},
  \bibnamefont{and} \bibinfo{author}{\bibfnamefont{J.~M.}
  \bibnamefont{Vaughan}}, \bibinfo{journal}{Phys. Rev. A}
  \textbf{\bibinfo{volume}{49}}, \bibinfo{pages}{3119 } (\bibinfo{year}{1994}).

\bibitem[{\citenamefont{Berkhout and Beijersbergen}(2008)}]{Berkhout:2008ty}
\bibinfo{author}{\bibfnamefont{G.~C.~G.} \bibnamefont{Berkhout}}
  \bibnamefont{and} \bibinfo{author}{\bibfnamefont{M.~W.}
  \bibnamefont{Beijersbergen}}, \bibinfo{journal}{Phys. Rev. Lett.}
  \textbf{\bibinfo{volume}{101}}, \bibinfo{pages}{100801}
  (\bibinfo{year}{2008}).

\bibitem[{\citenamefont{Paris and \v{R}eh\'a\v{c}ek}(2004)}]{Jarda:2004yk}
\bibinfo{editor}{\bibfnamefont{M.~G.~A.} \bibnamefont{Paris}} \bibnamefont{and}
  \bibinfo{editor}{\bibfnamefont{J.}~\bibnamefont{\v{R}eh\'a\v{c}ek}}, eds.,
  \emph{\bibinfo{title}{Quantum State Estimation}}, vol. \bibinfo{volume}{649}
  of \emph{\bibinfo{series}{Lecture Notes in Physics}}
  (\bibinfo{publisher}{Springer}, \bibinfo{address}{Heidelberg},
  \bibinfo{year}{2004}).

\bibitem[{\citenamefont{Lvovsky and Raymer}(2009)}]{Lvovsky:2009zk}
\bibinfo{author}{\bibfnamefont{A.~I.} \bibnamefont{Lvovsky}} \bibnamefont{and}
  \bibinfo{author}{\bibfnamefont{M.~G.} \bibnamefont{Raymer}},
  \bibinfo{journal}{Rev. Mod. Phys.} \textbf{\bibinfo{volume}{81}},
  \bibinfo{pages}{299} (\bibinfo{year}{2009}).


\bibitem[{\citenamefont{Rehacek et~al.}(2006)\citenamefont{
  \v{R}eh\'a\v{c}ek, Bouchal, {\v{C}}elechovsk\'{y}, Hradil and
  S{\'{a}}nchez-Soto}}]{Rehacek:2008ss}
\bibinfo{author}{\bibinfo{author}{\bibfnamefont{J.}~\bibnamefont{\v{R}eh\'a\v{c}ek}},
  \bibinfo{author}{\bibfnamefont{Z.}~\bibnamefont{Bouchal}},
  \bibinfo{author}{\bibfnamefont{R.}~\bibnamefont{{\v{C}}elechovsk\'{y}}},
\bibfnamefont{Z.}~\bibnamefont{Hradil}},
  \bibnamefont{and} \bibinfo{author}{\bibfnamefont{L.~L.}
  \bibnamefont{S{\'{a}}nchez-Soto}}, \bibinfo{journal}{Phys. Rev. A}
  \textbf{\bibinfo{volume}{77}}, \bibinfo{pages}{032110}
  (\bibinfo{year}{2008}).

\bibitem[{\citenamefont{Hradil et~al.}(2006)\citenamefont{Hradil,
  \v{R}eh\'a\v{c}ek, Bouchal, {\v{C}}elechovsk\'{y}, and
  S{\'{a}}nchez-Soto}}]{Hradil:2006bc}
\bibinfo{author}{\bibfnamefont{Z.}~\bibnamefont{Hradil}},
  \bibinfo{author}{\bibfnamefont{J.}~\bibnamefont{\v{R}eh\'a\v{c}ek}},
  \bibinfo{author}{\bibfnamefont{Z.}~\bibnamefont{Bouchal}},
  \bibinfo{author}{\bibfnamefont{R.}~\bibnamefont{{\v{C}}elechovsk\'{y}}},
  \bibnamefont{and} \bibinfo{author}{\bibfnamefont{L.~L.}
  \bibnamefont{S{\'{a}}nchez-Soto}}, \bibinfo{journal}{Phys. Rev. Lett.}
  \textbf{\bibinfo{volume}{97}}, \bibinfo{pages}{243601}
  (\bibinfo{year}{2006}).

\bibitem[{\citenamefont{Asorey et~al.}(2007)\citenamefont{
  Asorey, Facchi, Man'ko, Marmo, Pascazio and Sudarshan}}]{saverio}
\bibinfo{author}{\bibfnamefont{M.}~\bibnamefont{Asorey}},
  \bibinfo{author}{\bibfnamefont{P.}~\bibnamefont{Facchi}},
  \bibinfo{author}{\bibfnamefont{V.~I.}~\bibnamefont{Man'ko}},
\bibinfo{author}{\bibfnamefont{G.}~\bibnamefont{Marmo}},
\bibinfo{author}{\bibfnamefont{S.}~\bibnamefont{Pascazio}},
  \bibnamefont{and} \bibinfo{author}{\bibfnamefont{E.~G.~C.}
  \bibnamefont{Sudarshan}}, \bibinfo{journal}{Phys. Rev. A}
  \textbf{\bibinfo{volume}{76}}, \bibinfo{pages}{012117}
  (\bibinfo{year}{2007}).


\bibitem[{\citenamefont{Rigas et~al.}(2008)\citenamefont{Rigas,
  S{\'{a}}nchez-Soto, Klimov, \v{R}eh\'a\v{c}ek, and Hradil}}]{Rigas:2008ac}
\bibinfo{author}{\bibfnamefont{I.}~\bibnamefont{Rigas}},
  \bibinfo{author}{\bibfnamefont{L.~L.} \bibnamefont{S{\'{a}}nchez-Soto}},
  \bibinfo{author}{\bibfnamefont{A.~B.} \bibnamefont{Klimov}},
  \bibinfo{author}{\bibfnamefont{J.}~\bibnamefont{\v{R}eh\'a\v{c}ek}},
  \bibnamefont{and} \bibinfo{author}{\bibfnamefont{Z.}~\bibnamefont{Hradil}},
  \bibinfo{journal}{Phys. Rev. A} \textbf{\bibinfo{volume}{78}},
  \bibinfo{pages}{060101 (R)} (\bibinfo{year}{2008}).
\end{thebibliography}

\end{document}